\documentclass[aps,prd,twocolumn,groupedaddress,showpacs]{revtex4}
\usepackage{graphicx}
\usepackage{dcolumn}
\usepackage{bm}
\usepackage{amsmath}
\usepackage{epstopdf}
\usepackage{amsfonts}
\usepackage{amssymb}%

\usepackage[colorlinks=true,citecolor=blue,urlcolor=magenta,breaklinks]{hyperref}

\usepackage{natbib}

\providecommand{\U}[1]{\protect\rule{.1in}{.1in}}

\newcommand{\be}{\begin{equation}}
\newcommand{\en}{\end{equation}}
\newcommand{\bea}{\begin{eqnarray}}
\newcommand{\ena}{\end{eqnarray}}

\begin{document}

\title{Gravitational effects of condensed dark matter on strange stars}

\author{Grigorios Panotopoulos and Il\'\i dio Lopes}
\email{grigorios.panotopoulos@tecnico.ulisboa.pt, ilidio.lopes@tecnico.ulisboa.pt}
\affiliation{CENTRA, Instituto Superior T{\'e}cnico,\\ Universidade de Lisboa,
Av. Rovisco Pa{\'i}s 1, Lisboa, Portugal}

\date{\today}

\begin{abstract}
In the present work we study the gravitational effects of condensed dark matter on strange stars. We consider self-interacting dark matter particles with properties consistent with current observational constraints, and dark matter inside
the star is modelled as a Bose-Einstein condensate. We integrate numerically the Tolman-Oppenheimer-Volkoff equations in the two-fluid formalism assuming that strange stars are made of up to 4 per cent of dark matter.
It is shown that for a mass of the dark matter particles in the range $50 MeV-160 MeV$
strange stars are characterized by a maximum mass and radius similar to the ones found
for neutron stars.
\end{abstract}

\maketitle

\section{Introduction}

The $\Lambda$ Cold Dark Matter model ($\Lambda$CDM) that describes the formation of structure of the Universe from stars to galaxy clusters has been quite successful in explaining the observational characteristics of the present Universe. The success of the $\Lambda$CDM model is mainly due to that fact  
that all matter in the Universe is made of ordinary Standard Model particles and an unknown type of particles known as dark matter (DM). This new type of particle corresponds to 90\% of all matter in the present Universe. The discovery of DM  was made in 1933  by Zwicky when he was studying the dynamic properties of the Coma galaxy cluster~\cite{zwicky}. A few decades latter, Rubin and Ford arrived to similar conclusions about the existence of DM with optical studies of galaxies
like M31 ~\cite{rubin}. For a review on dark matter see e.g. \citep{munoz}. 

\smallskip

Many DM candidates have been recently proposed and studied by cosmologists and particle physicists alike \citep[e.g.,][]{taoso,2012RAA....12.1107T}
to explain or constrain the properties of DM~\citep{taoso,2012RAA....12.1107T}. Despite that, the origin and nature of DM still remains unknown. 
Indeed, the determination of the type of elementary particles that play the role of dark matter in the Universe is one of the current challenges 
of particle physics and modern cosmology. Among all the possible DM candidates, the most popular class consists of the Weakly Interacting Massive Particles (WIMPs). 
Like photons and neutrinos, these too are thermal relics from the Big-Bang. Initially the temperature of the Universe was high enough 
to maintain the  $\chi$  DM particles in equilibrium with the rest of the particles of the Standard Model. However, as the Universe cools down due to its expansion, at a certain point the annihilation rate of $\chi$ particles drops below the Hubble expansion rate of the Universe. When this happens the abundance of $\chi$ particles  freezes-out since the DM particles can no longer annihilate.  As a consequence their abundance remains constant ever since. Accordingly, today's DM relic density is given by \cite{SUSYDM} $ \Omega_{\chi} h^2 = {3 \times 10^{-27} cm^3 s^{-1}}/{\langle \sigma v \rangle_\chi} $ where $h$ is related to the Hubble constant $H_0=100 \: h (km s^{-1})/(Mpc)$. Since the DM particles do not have neither strong nor electromagnetic interactions, the WIMPs annihilation cross section typically has a value $\langle \sigma v \rangle_\chi = 3 \times 10^{-26} cm^3/s$ \cite{munoz}, and thus reproduces the current observed DM abundance $\Omega_\chi h^2=0.1198 \pm 0.0015$ \cite{wmap,planck2015}.

\smallskip 
 
The dark matter particles are usually assumed to be collisionless. However in \cite{self-interacting} the authors introduced the idea that
dark matter may have self interaction in order to alleviate some apparent conflicts between the collisionless cold dark mater paradigm and astrophysical
observations. It was found in \cite{self-interacting} that the appropriate range for the strength of self-interaction has to be
\begin{equation}
0.45 \frac{cm^2}{g} < \frac{\sigma_\chi}{m_\chi} < 450 \frac{cm^2}{g}
\end{equation}
where $m_\chi$ is the mass of the dark matter particles, and $\sigma_\chi$ the self interaction cross section of dark matter. Current limits on the strength
of the dark matter self interaction read \cite{bullet1,bullet2,review}
\begin{equation}
1.75 \times 10^{-4} \frac{cm^2}{g} < \frac{\sigma_\chi}{m_\chi} < (1-2) \frac{cm^2}{g}
\end{equation}

\smallskip

Compact objects \cite{textbook}, such as white dwarfs and neutron stars, are the final fate of stars.
The degeneracy pressure provided by the Fermi gas
balances the gravitational force, and the star finds a stable configuration. In white dwarfs the Fermi gas consists of
electrons, while in neutron stars the required pressure is provided by neutrons. Recently a new class of compact objects has
been postulated to exist due to some observed super-luminous supernovae \cite{superluminous}, which occur in about one out of
every 1000 supernovae explosions, and are more than 100 times brighter than normal supernovae. One plausible explanation is that neutrons are
further compressed so that a new object made of de-confined quarks is formed. This new compact object is called a "strange star", and since it is a much 
more stable configuration compared to a neutron star it could explain the origin of the huge amount of energy released in super-luminous
supernovae \cite{strangestars}. Compact objects, due to their unique properties, comprise an excellent natural laboratory to study and perhaps
constrain either modifications of gravity or physics beyond the Standard Model.

\smallskip

Even if dark matter does not interact directly with normal matter, it can have significant gravitational effects on stellar
objects \cite{massiveNS,chinos,admixed}. In \cite{admixed} the authors studied the gravitational effects of fermionic dark matter
on strange stars. However, since as of today the spin of the dark matter particles remains unknown, one could consider the
bosonic dark matter scenario~\cite{britoetal1,britoetal2}. As first pointed out
by Bose \cite{bose} and later expanded by Einstein \cite{einstein1,einstein2}, if the temperature of a quantum boson gas is low enough
or the number density of particles is large enough, a Bose-Einstein Condensate (BEC) is formed. The authors of \cite{darkstars}
modeled dark matter inside the star as a BEC with an equation of state of the form
\begin{equation}
p_\chi = \frac{2 \pi \hbar^2 l}{m_\chi^3} \rho_\chi^2
\end{equation}
where $\hbar$ is the reduced Planck constant, and $l$ is the scattering length and determines the self interaction cross section of
dark matter, $\sigma_\chi = 4 \pi l^2$ \cite{chinos,darkstars}.

\smallskip

It is the aim of the present article to study the effects of bosonic condensed dark matter on properties of strange stars.
Our work is organized as follows: after this introduction, we present the theoretical framework
in section two, and we constrain the scalar parameter space in the third section. Finally we conclude in section four.
We work in units in which the speed
of light in vacuum $c$ and the reduced Planck mass $\hbar$ are set equal to unity.
In these units all dimensionful quantities are measured in GeV, and we make use of the conversion rules
$1 m = 5.068 \times 10^{15} GeV^{-1}$ and $1 kg = 5.610 \times 10^{26} GeV$ \cite{guth}.

\section{The interior problem of relativistic stars}

We briefly review relativistic stars in General Relativity (GR). The starting point is Einstein's
field equations without a cosmological constant
\be
G_{\mu \nu} = R_{\mu \nu}-\frac{1}{2} R g_{\mu \nu}  = 8 \pi T_{\mu \nu}
\en
where we have set Newton's constant equal to unity, $G=1$, and in the exterior problem the matter energy momentum
tensor vanishes. For matter we assume a perfect fluid with pressure $p$, energy density $\rho$ and an equation of state
$p(\rho)$, while the energy momentum trace is given by $T=-\rho+3p$.
For the metric in the case of static spherically symmetric spacetimes we consider the following ansatz
\be
ds^2 = -f(r) dt^2 + g(r) dr^2 + r^2 d \Omega^2
\en
with two unknown functions of the radial distance $f(r), g(r)$. For the exterior problem one obtains the well-known solution
\be
f(r) = g(r)^{-1} = 1-\frac{2 M}{r}
\en
where $M$ is the mass of the star.
For the interior solution we introduce the function $m(r)$ instead of the function $g(r)$ defined as follows
\be
g(r)^{-1} = 1-\frac{2 m(r)}{r}
\en
so that upon matching the two solutions at the surface of the star we obtain $m(R)=M$, where $R$ is the radius of the star.
The Tolman-Oppenheimer-Volkoff (TOV) equations for the interior solution of a relativistic star with a
vanishing cosmological constant read \cite{TOV}
\bea
m'(r) & = & 4 \pi r^2 \rho(r) \\
p'(r) & = & - (p(r)+\rho(r)) \: \frac{m(r)+4 \pi p(r) r^3}{r^2 (1-\frac{2 m(r)}{r})}
\ena
where the prime denotes differentiation with respect to r, and the equations are to be integrated with the initial conditions
$m(r=0)=0$ and $p(r=0)=p_c$, where $p_c$ is
the central pressure. The radius of the star is determined requiring that the pressure vanishes at the surface,
$p(R) = 0$, and the mass of the star is then given by $M=m(R)$.

\section{Effect of condensed dark matter on strange stars}

Now let us assume that the star consists of two fluids, namely strange matter (de-confined quarks) and dark matter with only gravitational interaction between them, and equations of state $p_s(\rho_s)$, $p_\chi(\rho_\chi)$ respectively. The total pressure and the total energy density of the system are given by $p=p_s+p_\chi$ and $\rho=\rho_s+\rho_\chi$ respectively.
Since the energy momentum tensor of each fluid is separately conserved, the TOV equations in the two-fluid formalism for the interior solution
of a relativistic star with a vanishing cosmological constant read \cite{2fluid1,2fluid2}
\bea
m'(r) & = & 4 \pi r^2 \rho(r) \\
p_s'(r) & = & - (p_s(r)+\rho_s(r)) \: \frac{m(r)+4 \pi p(r) r^3}{r^2 (1-\frac{2 m(r)}{r})} \\
p_\chi'(r) & = & - (p_\chi(r)+\rho_\chi(r)) \: \frac{m(r)+4 \pi p(r) r^3}{r^2 (1-\frac{2 m(r)}{r})}
\ena
In this case in order to integrate the TOV equations we need to specify the central values both for normal matter and for
dark matter $p_s(0)$ and $p_\chi(0)$ respectively. So in the following we show the mass-radius diagram for a certain value of the constant $K=2 \pi l/m_\chi^3$ and for fixed dark matter fraction
\be
\epsilon = \frac{p_\chi(0)}{p_s(0)+p_\chi(0)}
\en
and we consider four cases, namely $\epsilon=0.02, 0.035, 0.05, 0.09$. We have chosen these values in agreement with the current dark matter constraints 
obtained from stars like the Sun. Actually, as shown by several authors, even smaller amounts of DM (as a percentage of the total mass of the star) can have a quite visible impact on the structure of these stars~\citep{2002PhRvL..88o1303L,ilidio2,ilidio5}. 
As we discuss in this work even such small amounts of DM can change the $M-R$ relation of neutron stars.

\smallskip

For strange matter we shall consider the simplest equation of state corresponding to a relativistic gas of de-confined quarks,
known also as the MIT bag model \cite{bagmodel}
\be
p_s = \frac{1}{3} (\rho_s - 4B)
\en
and the bag constant has been taken to be $B=(148 MeV)^4$ \cite{Bvalue}.
We mention in passing that for refinements of the bag model the reader could consult e.g. \cite{refine}, and for the present state-of-the-art
the recent paper \cite{art}. Strange stars with the above equations of state have been
investigated in \cite{prototype} for negative and positive values of the cosmological constant. There it was shown that the observed value of the
cosmological constant $\Lambda \sim (10^{-33} eV)^2$ is too small to have an observable effect. That is why in the present work we have taken the cosmological constant to be zero.

\begin{figure}[ht!]
\centering
\includegraphics[scale=0.70]{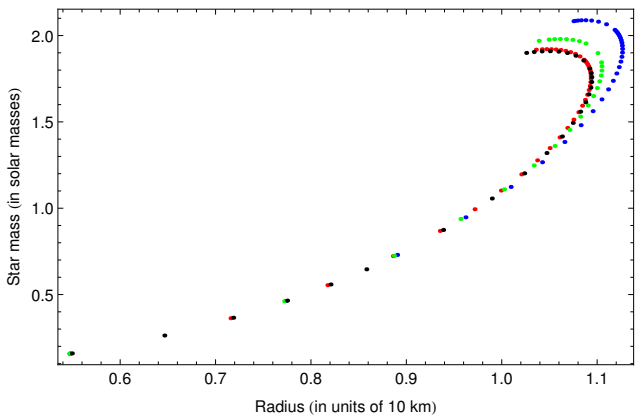}
\caption{Mass-radius diagram for $K=4/B$ and three different dark matter fractions, $\epsilon=0.02$ (red), $\epsilon=0.05$ (green) and $\epsilon=0.09$ (blue). The standard diagram without dark matter (black) is also shown for comparison.}
\label{fig:1} 	
\end{figure}

\begin{figure}[ht!]
\centering
\includegraphics[scale=0.70]{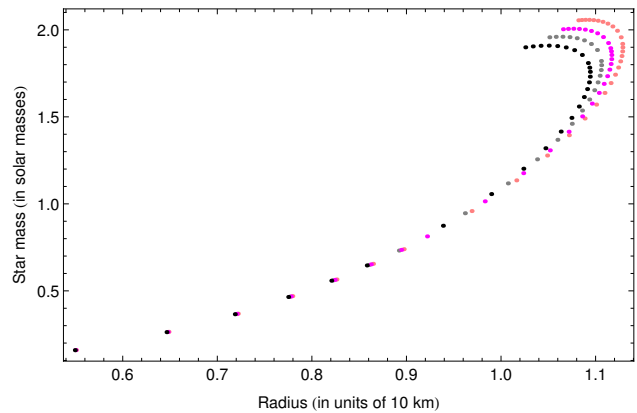}
\caption{Mass-radius diagram for $K=150/B$ and three different dark matter fractions, $\epsilon=0.02$ (gray), $\epsilon=0.035$ (magenta) and $\epsilon=0.05$ (pink). The standard diagram without dark matter (black) is also shown for comparison.}
\label{fig:1} 	
\end{figure}

\begin{figure}[ht!]
\centering
\includegraphics[scale=0.80]{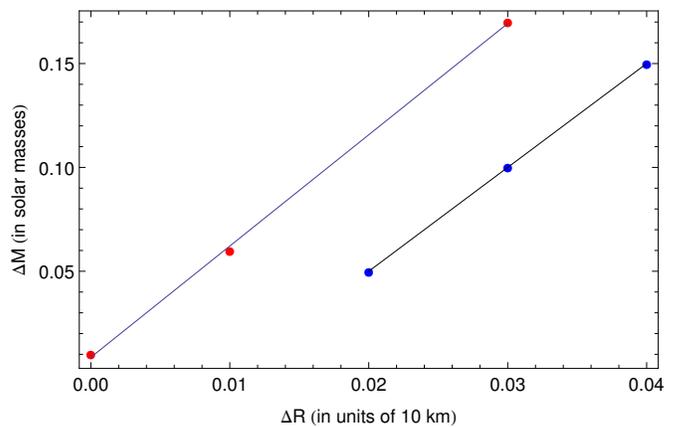}
\caption{Mass difference compared to standard results (without dark matter) versus radius difference for $KB=4$ (red) and $KB=150$ (blue).}
\label{fig:1} 	
\end{figure}

For the condensed dark matter we shall consider the equation of state obtained in \cite{darkstars}, namely $P_\chi=K \rho_\chi^2$,
where the constant $K=2 \pi l/m_\chi^3$ is given in terms of the mass of the dark matter particles $m_\chi$ and the scattering length $l$.
In a dilute and cold gas only the binary collisions at low energy are relevant, and these collisions are characterized by the s-wave scattering length
$l$ independently of the form of the two-body potential \cite{darkstars}. Therefore we can consider a short range repulsive delta-potential of the form
\begin{equation}
V(\vec{r}_1-\vec{r}_2) = \frac{4 \pi l}{m_\chi} \delta^{(3)}(\vec{r}_1-\vec{r}_2)
\end{equation}
which implies a dark matter self interaction cross section of the form $\sigma_\chi=4 \pi l^2$ \cite{chinos,darkstars}. Following previous studies
we fix the scattering length to be $l=1 fm$ \cite{chinos,darkstars}, and for $\sigma_\chi/m_\chi$ we apply the bounds discussed in the Introduction
\begin{equation}
0.45 \frac{cm^2}{g} < \frac{\sigma_\chi}{m_\chi} < 1.5 \frac{cm^2}{g}
\end{equation}
which then implies the following range for the mass of the dark matter particle
\begin{equation}
0.05 GeV < m_\chi < 0.16 GeV
\end{equation}
and thus for the constant $K$
\begin{equation}
\frac{4}{B} < K < \frac{150}{B}
\end{equation}
where now the constant $K$ is given in units of the bag constant.

\smallskip

In Fig.~1 we show the mass-radius diagram for the lowest allowed value of K, namely $K=4/B$, and three different dark matter fractions, while in
Fig.~2 we show the mass-radius diagram for the maximum value of K, namely $K=150/B$, and three different dark matter fractions. For comparison the standard
diagram without dark matter is also shown. Our numerical results show that dark matter only affects the more massive stars. The maximum radius $R_{max}$ as well as the maximum star mass $M_{max}$ for each case can be seen in Table~1. The dark matter mass fraction is also shown. For a given constant $K$, $R_{max}$ and $M_{max}$ increase with the dark matter fraction. Therefore, if strange stars do exist, and if they accumulate dark matter, our findings set certain limits on the properties of these hypothetical objects.

\begin{table}[ht]
\caption{Maximum stellar mass (in unities of $M_\odot$) and radius (in units of $10\; km$) of a strange star with different fractions of dark matter mass $M_{dm}/M$ .}
\centering
\begin{tabular}{c | c c c c c}
\hline	
$m_\chi$ & K & $\epsilon$ & $R_{max}$ & $M_{max}$ & $M_{dm}$ \\
GeV    & $B^{-1}$\footnote{In our calculation
$B=(148 MeV)^4$.}

  & $-$   &  10 km & $M_{\odot}$ & $M$   \\
\hline
{\bf no DM}\footnote{Note that the strange star (standard model) without dark mater has a  $R_{max}=1.09$ and $M_{max}=1.91$.}    & 0  & 0 & 1.09 & 1.91 & 0  \\
0.16 & 4 & 0.02 & 1.09 & 1.92 & 0.0146 \\
0.16 & 4 & 0.05 & 1.10 & 1.97 & 0.0271 \\
0.16 & 4 & 0.09 & 1.12 & 2.08 & 0.0436 \\
0.05 & 150 & 0.02 & 1.11 & 1.96 & 0.0032 \\
0.05 & 150 & 0.035 & 1.12 & 2.01 & 0.0044 \\
0.05 & 150 & 0.05 & 1.13 & 2.06 & 0.0057
\end{tabular}
\label{table:First set}
\end{table}

Finally, in Fig.~3 we show the mass difference $\Delta M = M(\epsilon)-M_{st}$ versus the radius difference $\Delta R=R(\epsilon)-R_{st}$, where $M(\epsilon)$
is the maximum star mass for each case shown in Table 1, while $R(\epsilon)$ is the maximum star radius for the same case, and $M_{st}=1.91$, $R_{st}=10.9\;{km}$
are the standard values without dark matter.

\section{Conclusions}

In the present article we have studied the gravitational effects of condensed dark matter on strange stars, considering self-interacting dark matter particles with properties consistent with current observational constraints.
Strange stars are hypothetical
compact objects that are supposed to be much more stable than neutron stars, and thus could explain the super luminous supernovae.
We have considered the star interior problem assuming spherical symmetry, and we have solved numerically the Tolman-Oppenheimer-Volkoff equations in the two-fluid formalism. For strange matter we have assumed the simplest version of the MIT bag model (radiation plus the bag constant), while if dark matter is modeled inside the star as a BEC,
it can be described by a polytrope
equation of state $P_\chi=(2 \pi l/m_\chi^3) \rho_\chi^2$, where $m_\chi$ is the mass of the dark matter particles, and $l$ is the scattering length determining the
self interaction cross section of dark matter in the form $\sigma_\chi=4 \pi l^2$. We show the mass-radius diagram assuming
that strange stars are made of up to 4 per cent of dark matter, and for a mass of the dark matter particles in the range $50 MeV-160 MeV$.
We conclude that if strange stars do exist, and if they accumulate dark matter, our findings limit in a certain way the radius and the mass of these compact objects.


\begin{acknowledgments}
The authors wish to thank the anonymous reviewer for his/her suggestions. This work was supported 
by "Funda{\c c}{\~a}o para a Ci{\^e}ncia e Tecnologia".
\end{acknowledgments}


\end{document}